\documentclass[12pt,english]{article}
\usepackage{lmodern}
\usepackage{lmodern}
\usepackage[T1]{fontenc}
\usepackage[cp1252]{inputenc}
\usepackage{geometry}
\usepackage{color}
\usepackage{babel}
\usepackage{amsmath}
\usepackage{amsthm}
\usepackage{amssymb}
\usepackage{setspace}
\usepackage[authoryear,comma, longnamesfirst]{natbib}
\usepackage{microtype}
\usepackage[unicode=true,
 bookmarks=false,
 breaklinks=false,pdfborder={0 0 1},backref=section,colorlinks=true]
 {hyperref}
\hypersetup{pdftitle={"DRUM"},
 pdfauthor={"KashaevAguiar"},
 pdfnewwindow=true,pdfstartview=FitH,urlcolor=blue!90!red!45!black,citecolor=blue!90!red!45!black,linkcolor=red!90!black}

\makeatletter
\theoremstyle{plain}
\newtheorem{thm}{\protect\theoremname}
\theoremstyle{definition}

\theoremstyle{definition}
\newtheorem{example}[thm]{\protect\examplename}
\theoremstyle{plain}

\usepackage{amsfonts}
\usepackage{dsfont}\usepackage{mathrsfs}\usepackage{ushort}

\usepackage{titlesec}\usepackage{titling}
\usepackage{caption}
\usepackage{enumitem}\usepackage{booktabs}
\usepackage{tikz}
\usetikzlibrary{decorations.pathreplacing}
\usepackage{pstricks}\usepackage{pst-all}
\usepackage{pst-plot}
\usepackage{pst-node}\usepackage{pst-3dplot}\usepackage{sgamevar}
\usepackage{subcaption}
\usepackage{lscape}
\usepackage{pifont}
\usepackage{longtable}
\usepackage{algorithmic}
\usepackage{pdflscape}
\usepackage{rotating}
\usepackage{float}

\setenumerate{label=\small(\roman*)}
\DeclareCaptionFont{fancy}{\bfseries\sffamily}
\gamemathtrue
\allowdisplaybreaks

\providecommand{\psreset}{\psset{%
		linewidth=0.3pt,linestyle=solid,linecolor=black,
		dotsize=2.5pt,dotsep=2.5pt,arrowsize=4pt,
		fillstyle=none,fillcolor=white,
		showpoints=false,arrows=-,linearc=0,framearc=0,
		hatchsep=2pt,hatchwidth=0.2pt,nodesep=4pt,opacity=1}
	\psset{gridcolor=black!60, subgridcolor=black!30}
}

\psreset

\usepackage{graphicx}
\graphicspath{ {figures/} }

\titleformat{\section}[block]{\centering\large\bfseries\sffamily}{\thesection.}{0.5em}{}
\titleformat{\subsection}[block]{\flushleft\bfseries}{\thesubsection.}{0.5em}{}
\titleformat{\subsection}[block]{\flushleft\bfseries\sffamily}{\thesubsection.}{0.5em}{}
\titleformat{\subsubsection}[runin]{\normalsize\bfseries\sffamily}{\bfseries\upshape\sffamily\thesubsubsection.}{0.5em}{}[.--\:]
\renewcommand{\thesubsubsection}{\arabic{section}.\arabic{subsection}.\arabic{subsubsection}}
\titlespacing{\section}{0ex}{10ex}{5ex}
\titlespacing{\subsection}{0in}{6ex}{3ex}
\titlespacing{\subsubsection}{0mm}{2ex}{0.5em}
\pretitle{\begin{center}\LARGE\bfseries\sffamily}
\posttitle{\par\end{center}\vskip 0.5em}
\preauthor{\begin{center} \large \lineskip 0.5em\begin{tabular}[t]{c}}
\postauthor{\end{tabular}\par\end{center}}
\predate{\begin{center}\small}
\postdate{\par\end{center}}
\providecommand{\abstitle}[1]{{\par\vspace*{2ex}\small\bfseries\sffamily #1}\hspace*{1ex}}
\renewenvironment{abstract}%
{\begin{center}\begin{minipage}{0.8\linewidth}%
			\abstitle{Abstract}\small}%
		{\end{minipage}\end{center}\vfill\clearpage}

\DeclareMathOperator*{\argmax}{arg\,max}

\providecommand{\Char}[1]{\mathds{1}\left(\,#1\,\right)}
\providecommand{\Real}{{\mathds{R}}}

\providecommand{\tr}{^{\prime}}
\providecommand{\as}{\ensuremath{\mathrm{a.s.}}}
\providecommand{\rand}[1]{\mathbf{#1}}

\providecommand{\abs}[1]{\left\lvert#1\right\rvert}

\theoremstyle{remark}
  \theoremstyle{plain}
  \newtheorem{lemma}{\protect\lemmaname}\theoremstyle{definition}
    \newtheorem{proposition}{\protect\propositionname}\theoremstyle{definition}
  \newtheorem{definition}{\protect\definitionname}\theoremstyle{plain}
\newtheorem{theorem}{\protect\theoremname}\theoremstyle{plain}
  \theoremstyle{definition}
  \providecommand{\assumptionname}{Assumption}

  \providecommand{\definitionname}{Definition}
  \providecommand{\lemmaname}{Lemma}
  \providecommand{\propositionname}{Proposition}
  \providecommand{\remarkname}{Remark}
\providecommand{\corollaryname}{Corollary}
\providecommand{\theoremname}{Theorem}
\providecommand{\examplename}{Example}

\makeatother

\providecommand{\definitionname}{Definition}
\providecommand{\examplename}{Example}
\providecommand{\lemmaname}{Lemma}
\providecommand{\theoremname}{Theorem}

\begin{document}
\title{Nonparametric Analysis of Dynamic Random Utility Models\thanks{The ``\textcircled{r}'' symbol indicates that the authors' names are in certified random order, as described by \citet{ray2018certified}. We thank Roy Allen, Adam Dominiak, David Freeman, Matt Kovach, Krishna Pendakur, and Tomasz Strzalecki for useful discussions and encouragement.}}
\author{ 
	Nail Kashaev \textcircled{r}
	Victor H. Aguiar\thanks{Kashaev: Department of Economics, University of Western Ontario; \href{mailto:nkashaev@uwo.ca}{nkashaev@uwo.ca}. Aguiar: Department of Economics, University of Western Ontario; \href{mailto:vaguiar@uwo.ca}{vaguiar@uwo.ca}.}
	}
\date{April, 2022}
\maketitle

\begin{abstract}
We study a dynamic generalization of stochastic rationality in consumer behavior, the Dynamic Random Utility Model (DRUM). Under DRUM, a consumer draws a utility function from a stochastic utility process and maximizes this utility subject to her budget constraint in each time period. Utility is random, with unrestricted correlation across time periods and unrestricted heterogeneity in a cross-section. We provide a revealed preference characterization of DRUM  when we observe a panel of choices from budgets. This characterization is amenable to statistical testing. Our result unifies Afriat's \citeyearpar{afriat1967construction} theorem that works with time-series data and the static random utility framework of McFadden-Richter \citeyearpar{mcfadden1990stochastic} that works with cross-sections of choice. 


\end{abstract}

\section{Introduction}
One key question in economics is whether consumer behavior is rational. Traditional definitions of rationality are effectively equivalent to maximizing a utility function that is fixed in time. Here, we study a notion of rationality in consumer behavior that is stochastic and dynamic -- Dynamic Random Utility Model (DRUM). This model allows for unrestricted heterogeneity in preferences across consumers and does not restrict the correlation of preferences of the same consumer across time. Under DRUM, each consumer in each period of time maximizes the realized utility from a stochastic utility process subject to a budget constraint. 
\par 
Theoretically, we provide a synthesis of the two main paradigms of nonparametric demand analysis, the Afriat's framework and the McFadden-Richter framework. The framework developed in Afriat \citeyearpar{afriat1967construction} requires us to observe a \emph{time series} of choices and budgets of a given consumer. The Afriat's theorem characterizes utility maximization under the assumption that a consumer maximizes the same utility function each time period (i.e., preferences do not change with time). When this assumption about the utility stochastic process being constant over time is relaxed, there are no empirical implications when observing only a time-series of choices. The framework developed in McFadden-Richter \citeyearpar{mcfadden1990stochastic}, called random utility model (RUM), instead requires us only to observe a \emph{cross-section} of choices and budgets of a population of consumers. There is no time-dimension in RUM. One can ignore the panel structure and study a \emph{slice} of the panel in a given time period (i.e., a slice is the (marginal) probability of choice over budgets in a given time period).  Unfortunately, this approach ignores the potential correlation among time of the utility stochastic process. As a result, there are certain panels of choices over budgets that when sliced are consistent with RUM, but they cannot be rationalized by DRUM. In other words, ignoring the time-series dimension of choice may lead to false positives when testing RUM. In this paper, we consider a richer primitive that unifies these two frameworks. This unification is advantageous because it (i) provides more informative bounds on counterfactual choice due to the richer variation in the panel of choices; (ii) provides a theoretical justification for slicing choices and using the RUM framework; and (iii) clarifies the role of the constant preferences across time assumption in the Afriat's framework that allows to test rationality using only time-series of choices. Fortunately, our primitive with a longitudinal level of variation is readily available in many consumption surveys, household scanner datasets, and experimental dataset as documented in Aguiar and Kashaev \citeyearpar{AK2021}.
\par
Note that in practice, panels of choices are often pooled in the time dimension to create a cross-section with sufficient variation of budgets \citep{deb2017revealed,kitamura2018nonparametric}. We show this approach could lead to false rejections of DRUM, due to ignoring the time labels of budgets. 
\par
Empirically, DRUM is needed because the notion of static utility maximization in the Afriat's framework is under scrutiny. Experimental and field evidence show utility maximization being violated in several domains.\footnote{For examples in household consumption see \citet{echenique2011money,dean2016measuring} and in choices over portfolios over risk or uncertainty see \citet{choi2007revealing,choi2014more,ahn2014estimating}.} Moreover, there is evidence that some failures of the traditional static utility maximization model may be driven by the stringent assumption of the stability of preferences over time. For example, \citet{kurtz2019neural} document violations of static rationality due to variability of the neural computation of value in time. In consumption surveys, structural breaks in patterns of consumption is a well-documented phenomenon that cannot be accommodated by the standard static utility maximization framework. \citet{cherchye2017household} provide evidence of structural breaks in dietary patterns within a year for an individual consumer, and provide a model of changing selves where utility changes deterministically in time.
\par 
In stark contrast with the Afriat's framework, RUM has found reasonable success explaining repeated cross-sections of household choices \citep{kawaguchi2017testing,kitamura2018nonparametric}. However, the McFadden-Richter's framework cannot take advantage of the longitudinal variation of choice that is available in many datasets \citep{im2021non}. By considering a richer primitive, we can at the same time relax the assumption of a stable utility function over time implicit in the Afriat's framework while providing a more informative test of stochastic utility  maximization than in the McFadden-Richter's framework. 
\par 
We provide a revealed preference characterization of DRUM when the longitudinal distribution of demand is observed for a finite collection of budgets in a finite time window. Notably, this characterization does not make any parametric restriction on (i) the form of utility functions, (ii) the correlation of utilities in time, and (iii) the heterogeneity of utility in the cross-section. This characterization lends itself to statistical testing and can be used for nonparametric counterfactual and welfare analysis that is robust to evolving and heterogeneous preferences.  
\par 
The DRUM framework is rich and extends well beyond the Afriat's and McFadden-Richter world. We cover as special cases: (i) consumption models of errors in the evaluation of utility \citep{kurtz2019neural}; (ii) dynamic random expected utility (defined in \citet{frick2019dynamic}) for choices over portfolios of securities as in \citet{polisson2020revealed}; (iii) static utility maximization in a population (without measurement error) \citep{AK2021}; (iv) dynamic utility maximization in a population\footnote{This requires a redefinition of price to be an effective price that includes an adjustment due to interests rates as described in \citet{AK2021}.} \citep{browning1989anonparametric,gauthier2018,AK2021}; (v) changing utility or multiple-selves models \citep{cherchye2017household}; and changing-taste modeled with a constant utility in time with an additive shock \citep{adams2015prices}. 
\par 
DRUM was first defined in \cite{straleckinotes} in an abstract domain for discrete choice. \citet{frick2019dynamic} provide an axiomatic characterization of it for a rich domain with decision trees and an expected utility restriction on the stochastic utility process. We provide the first characterization of DRUM for a consumer choice domain with limited observability on budgets without requiring any restriction on preferences. 
\par 
Recent interest in DRUM in finite abstract discrete choice space has provided partial characterizations of it when the primitive is the joint distribution of choices across time and with full menu variation. \citet{li2021axiomatization} provides an axiomatic characterization of DRUM for two time periods and full menu variation. \citet{chambers2021correlated} consider correlated choice which is the joint distribution of choice on a pair of menus, the choice may be made by a group instead of a single decision maker. Some versions of this model can be though as dynamic choice when the multiple selves of a decision maker are making decisions. However, the primitive in both \citet{li2021axiomatization} and \citet{chambers2021correlated} differs from ours in the general setup. Importantly, in our setup the domain of classical consumer choice  is endowed with a primitive order (i.e., the vector order), and  preference revelation respects that primitive order. Our DRUM will respect this primitive order and restrict the stochastic utility process to be monotone. Another difference is that we deal with a continuum of choices, and limited observability of menus and histories. Finally, \citet{li2021axiomatization} and \citet{chambers2021correlated} assume comprehensive menu variation that allows them to provide a characterization analogous to \citet{block1960random} exploiting the nested structure of menus under the set containment. In contrast to them, choice sets in our setup are not nested, so we cannot use the characterizations in \citet{li2021axiomatization} and \citet{chambers2021correlated}. 
\par 
\citet{AK2021} studies a panel setup as well but uses a first-order-conditions approach to deal with some forms of dynamic preferences. Mainly, they allow for measurement error, that can be mapped to trembling-hand or misperception-errors. However, their setup does not allow for changing utility beyond a changing discount factor or marginal utility of income.  \citet{im2021non} study the McFadden-Richter's framework and its inability to use a panel structure. However, they propose to check individual static rationality like in Afriat's framework as a potential solution. Here, we generalize the Afriat's framework to allow individual's utility to change over time, while exploiting the panel structure to obtain more empirical implications than in the McFadden-Richter framework. 
\par 
The paper is organized as follows, Section~\ref{sec: setup} introduces the setup. Section~\ref{sec: characterization of DRUM} provides a characterization of DRUM. In Section~\ref{sec: 2x2 case}, we study a simple setup with two time periods and two budgets in each time period. Section~\ref{sec:generalcasenecessity} provides computationally convenient testable implications of DRUM in the general setup. Section~\ref{sec: unification} provides a unification of the Afriat and McFadden-Richter's setup. Section~\ref{sec:pooling} provides a study of pooling a panel dataset when the dataset is consistent with DRUM. Section~\ref{sec: conclusion} concludes.  All proofs can be found in Appendix~\ref{app: proofs}.

\section{Setup}\label{sec: setup}
Let $X\subseteq \Real^K_{+}$ be the consumption space with finite $K\geq 2$ goods. We consider a time window $\mathcal{T}=\{1,\cdots,T\}$ with a finite terminal period $T\geq2$. In each time period $t\in \mathcal{T}$, we assume the existence of $J^t<\infty$ budgets
\[
B^t_{j}=\left\{y\in \Real^K_+\::\:p_{j,t}\tr y=w_{j,t}\right\}
\]
for all $j\in\mathcal{J}^t=\{1,\dots, J^t\}$, where $p_{j,t}\in \Real^K_+$ is the vector of prices and $w_{j,t}>0$ is the expenditure level. Let $J=\sum_{t\in \mathcal{T}}J^t$ denote the total number of budgets. 
\par
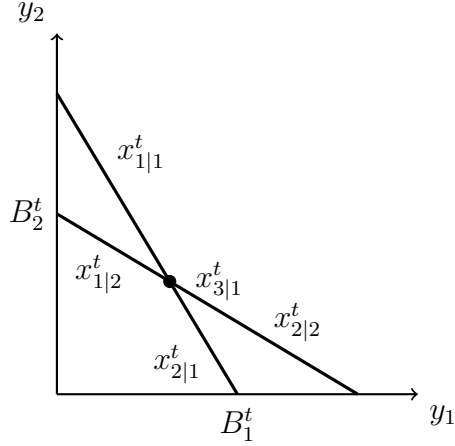
\begin{figure}
\begin{center}
\begin{tikzpicture}[scale=0.8]
\draw[thick,->] (0,0) -- (6,0) node[anchor=north west] {$y_1$};
\draw[thick,->] (0,0) -- (0,6) node[anchor=south east] {$y_2$};
\draw[very thick] (0,5) -- (3,0);
\draw[very thick] (5,0) -- (0,3);
\draw [fill=black] (1.875,1.875) circle[radius=.1];
\draw (2.7,1.875) node {$x^t_{3|1}$};
\draw (0.7,2) node {$x^t_{1|2}$};
\draw (2,0.5) node {$x^t_{2|1}$};
\draw (1.4,4) node {$x^t_{1|1}$};
\draw (4,1.2) node {$x^t_{2|2}$};
\draw (3,-0.5) node {$B^t_{1}$};
\draw (-0.5,3) node {$B^t_{2}$};
\end{tikzpicture}
\end{center}
\caption{Patches for the case with $K=2$ goods and $J_t=2$ budgets. The only intersection patch is $x^t_{3|1}$, which is the intersection of $B^t_{1}$ and $B^t_{2}$.}\label{fig: budgets_general}
\end{figure}

Following \citet{kitamura2018nonparametric} (henceforth KS), we introduce the notion of patches. For any $t\in\mathcal{T}$ and $j\in\mathcal{J}^t$, let $\{x^t_{i|j}\}$ be a finite partition of $B^t_{j}$, where each element of the partition is indexed by $i$.
\begin{definition} [Patches] For every $t\in\mathcal{T}$, let 
\[
\mathcal{X}^t=\bigcup_{j\in \mathcal{J}^t} \{x^t_{i|j}\}
\]
be the coarsest partition of $\bigcup_{j\in\mathcal{J}^t}B^t_{j}$ such that
\[
x^t_{i|j}\bigcap B^t_{j'}\in\{x^t_{i|j},\emptyset\}
\]
for any $j,j'\in \mathcal{J}^t$ and $i\in \mathcal{I}^t_j=\{1,\dots, I^t_{j}\}$, where $I^t_{j}$ is the cardinality of the partition $\{x^t_{i|j}\}$. The elements of $\mathcal{X}^t$ are called patches. If $x^t_{i|j}\subseteq B^t_{j'}$ for some $i$ and $j\neq j'$, then $x^t_{i|j}$ is called an intersection patch.
\end{definition}
By definition patches can only be strictly above, strictly below, or on budget planes. A typical patch belongs to one budget plane. However, intersection patches always belong to several budget planes. The case for $K=2$ goods and $J_t=2$ budgets is depicted in Figure~\ref{fig: budgets_general}.
Note that by definition $\{x^t_{i|j}\}$ is a partition of $B^t_{j}$ and $I^t_{j}$ is the number of patches that form budget $B^t_{j}$.
\par 
Define a budget path as a collection of indexes $\rand{j}=\{j_t\}_{t=1}^T$ such that $j_t\in\mathcal{J}^t$ for all $t$. Budget paths encode budgets they were faced by agents in different time periods. Let $\rand{J}$ be a set of observed budget paths. Given a budget path $\rand{j}\in \rand{J}$, a \emph{choice path} as the array of patches $x_{\rand{i}|\rand{j}}=\{x^t_{i_t|{j_t}}\}_{t\in\mathcal{T}}$ for some collections of indexes, $\rand{i}=\{i_t\}_{i=1}^T$ such that $i_t\in\mathcal{I}^t_{j_t}$ for all $t$. Similar to budget paths, choice paths encode choice of agents in a given sequence of budget sets she faced. The set of all possible choice path index sets $\rand{i}$ given the budget path $\rand{j}$ is denoted by $\rand{I}_\rand{j}$.
Let $\rho(x_{\rand{i}|\rand{j}})$ be the probability of observing a choice path $x_{\rand{i}|\rand{j}}$ given a budget path $\rand{j}$. That is, for a given $\rand{j}$, $\rho(x_{\rand{i}|\rand{j}})\geq 0$ and $\sum_{\rand{i}\in\rand{I}_{\rand{j}}}\rho(x_{\rand{i}|\rand{j}})=1$.  
\par
We assume that the longitudinal distribution of demand is known (can be consistently estimated). That is, the researcher observes a \emph{dynamic stochastic demand} 
\[
\rho=(\rho(x_{\rand{i}|\rand{j}}))_{\rand{j}\in\rand{J},\rand{i}\in\rand{I}_\rand{j}}.
\]
Some examples of our primitive are: (i) household longitudinal survey datasets (e.g., \textit{Encuesta de Presupuestos Familiares} in Spain); (ii) scanner datasets (e.g., Nielsen homescan data); and (iii) experimental datasets where subjects need to pick a point on the budget line several times (e.g., experiments on preferences over giving as in \citealp{porter2016love}). In survey datasets information about household purchases is usually collected several times a year (e.g., quarterly). For a given time period, budget variation across households is driven by spatial (e.g. regional) price variation \citep{AK2021}. Scanner datasets contain information about weekly purchases of consumers. Budget variation in this case is driven by price variation across stores in each time period \citep{gauthier2021}. In experimental settings, often, each subject faces at random a budget path drawn from the same set of budgets for all subjects. Since the number of subjects is usually much bigger than the number of budget paths, there are many subjects facing the same budget path.  
\par
We study the problem of whether a dynamic stochastic demand, $\rho$, can be rationalized by a dynamic random utility model (DRUM). To formally define DRUM we need some preliminaries. 
\par
Let $U$ denote the set of all continuous, strictly concave, and monotone utility functions that map $X$ to $\Real$. Also let $\mathcal{U}=\times_{t\in\mathcal{T}}U$ be the Cartesian product of $T$ repetitions of $U$.
\begin{definition} [DRUM] A dynamic stochastic demand $\rho$ is dynamically and stochastically rationalized by (or is consistent with) DRUM if there exists a probability measure over $\mathcal{U}$, $\mu$, such that
\[
\rho(x_{\rand{i}|\rand{j}})=\int \prod_{t\in \mathcal{T}} \Char{\argmax_{y\in B^t_{j}}u^t(y)\in x^t_{i_t|j_t}}d\mu(u)
\]
for all $\rand{j}\in\rand{J}$ and $\rand{i}\in\rand{I}_{\rand{j}}$, where $u=(u^t)_{t\in\mathcal{T}}$.
\end{definition}
We let $\rand{u}=(\rand{u}^t)_{t\in\mathcal{T}}$ summarize the stochastic utility process captured by $\mu$. DRUM implicitly imposes some exclusion restrictions on $\rand{u}$ that are analogous to the McFadden-Richter framework. Mainly, the distribution of $\rand{u}$ does not depend on the budget paths and does not depend on the alternatives in the consumption space. Nevertheless, and relaxing the implicit assumption in McFadden-Richter, $\rand{u}$ does not restrict the correlation between preferences across time, nor it restricts the preference heterogeneity in cross-sections. The Afriat's framework instead imposes a strict restriction that preferences are perfectly correlated across time (i.e. $\rand{u}^t=\rand{u}^s\:\as$ for all $t,s\in\mathcal{T}$), but does not restrict the preference heterogeneity across consumers. We formalize these connections in Section~\ref{sec: unification}. 

\section{Characterization of DRUM}\label{sec: characterization of DRUM}
Here we provide a characterization rationalizability by DRUM when $\rho$ is observed or estimable. The main result in this section will be an exact analogue of the McFadden-Richter and KS's results for RUM where there is no time variation. KS show that without loss of generality, we can work with patches rather than with actual consumption bundles. In particular, we can focus on ``representative'' elements of patches (e.g. geometric centers) and identify DRUM with a mixture over a finite number of preference profiles (linear orders) defined over the elements of $\mathcal{X}^t$. Let a preference profile be $\rand{r}=\{r_1,\cdots,r_T\}$, where $r_t$ is a linear order defined on $\mathcal{X}^t$. 
Given the preference profile $\rand{r}$, we can encode choices in different time periods and budgets in a vector $a_{\rand{r}}$ as 
\[
a_{\rand{r}}=\left(a_{\rand{r},\rand{i},\rand{j}}\right)_{\rand{j}\in\rand{J},\rand{i}\in\rand{I}_\rand{j}},
\]
with $a_{\rand{r},\rand{i},\rand{j}}=1$ if the patch $x^t_{i_t|j_t}$ is the best patch in $B^t_{j_t}$ according to $r_t$ for all $t\in\mathcal{T}$ and $a_{\rand{r},\rand{i},\rand{j}}=0$ otherwise. 
The set of dynamic rational preference profiles $\mathcal{R}$ is the set of all profiles of preferences $\rand{r}$ for which there exists $u_r=(u^t_{r})_{t\in\mathcal{T}}\in\mathcal{U}$ such that
\[
a_{\rand{r},\rand{i},\rand{j}}=1\quad\iff \quad\forall t\in\mathcal{T},\: \argmax_{x\in B_{j_t}}u^t_{r}(x)\subseteq x_{i_t|j_t}.
\]
\par 
We form matrix $A$ by stacking the column vectors $a_{\rand{r}}$ for all preference profiles in $\rand{r}\in\mathcal{R}$. The dimension of this matrix is $d_{\rho}\times \abs{\mathcal{R}}$, where $d_{\rho}$ is the length of vector $\rho$. This matrix will be used to provide a characterization of DRUM that is amenable to statistical testing. 
\par
The next axiom is the analogue of the McFadden-Richter axiom for (static) stochastic revealed preferences \citep{border2007introductory}.
\begin{definition} [Axiom of Dynamic Stochastic Revealed Preference, ADSRP]  
A dynamic stochastic demand $\rho$ satisfies ADSRP if for every finite sequence of pairs of budget and choice paths (including repetitions), $k$, $\{(\rand{i}_k,\rand{j}_k)\}$ such that $\rand{j}_k\in\rand{J}$ and $\rand{i}_k\in\rand{I}_{\rand{j}_k}$ 
\[
\sum_{k}\rho(x_{\rand{i}_k|\rand{j}_k})\leq \max_{\rand{r}\in\mathcal{R}}\sum_{k}a_{\rand{r},\rand{i}_k,\rand{j}_k}.
\]
\end{definition}

The next theorem provides a full characterization of DRUM.

\begin{theorem}\label{thm:main} 
The following are equivalent:
\begin{enumerate}
    \item The dynamic stochastic demand $\rho$ is dynamically rationalizable by DRUM.
    \item There exists $\nu\in \Delta^{|\mathcal{R}|-1}$ such that $\rho=A\nu$.
    \item There exists $\nu\in \Real^{|\mathcal{R}|}_{+}$ such that $\rho=A\nu$.
    \item The dynamic stochastic demand dataset $\rho$ satisfies the ADSRP. 
\end{enumerate}
\end{theorem}
The main part of the proof of Theorem~\ref{thm:main} is based on the fact that, without loss of generality, $\rho$ can be reduced to a demand that assigns mass only to the representative elements of patches (e.g., geometric centers) along a choice path. Then the equivalence of (i)-(iv) is analogous to proof for RUM in \citet{mcfadden1990stochastic,mcfadden2005revealed} and KS.  
\par 
Theorem~\ref{thm:main}(iii) is amenable to statistical testing using the tools developed in KS and the computational tools to compute matrix $A$ in \citet{smeulders2021nonparametric}. 
\par 
Next, we provide a simpler characterization of DRUM for a simple-setup. This will demonstrate that DRUM provides additional implications in longitudinal data than those in the McFadden-Richter's framework. 

\section{The Simple-Setup: $2$ time periods, $2$ budgets}\label{sec: 2x2 case}
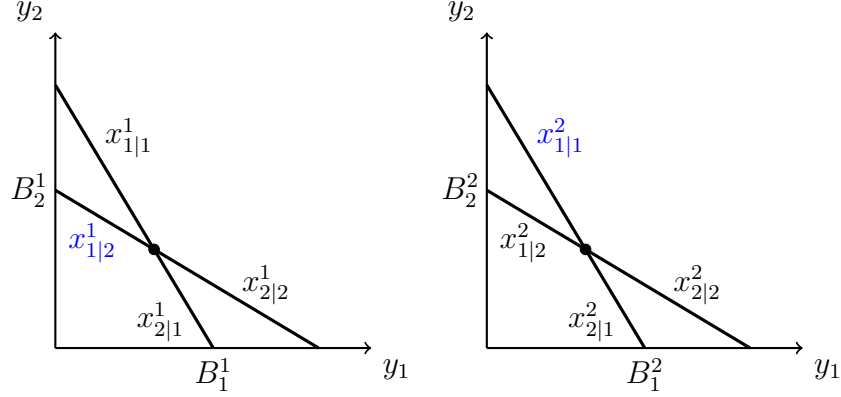
\begin{figure}[h]
\begin{centering}
\begin{tikzpicture}[scale=0.7] 
\draw[thick,->] (0,0) -- (6,0) node[anchor=north west] {$y_1$}; 
\draw[thick,->] (0,0) -- (0,6) node[anchor=south east] {$y_2$}; 
\draw[very thick] (0,5) -- (3,0); 
\draw[very thick] (5,0) -- (0,3); 
\draw [fill=black] (1.875,1.875) circle[radius=.1]; 
\draw (0.7,2) node [color=blue] {$x^1_{1|2}$}; \draw (2,0.5) node {$x^1_{2|1}$}; \draw (1.4,4) node {$x^1_{1|1}$}; \draw (4,1.2) node {$x^1_{2|2}$}; \draw (3,-0.5) node {$B^1_{1}$}; \draw (-0.5,3) node {$B^1_{2}$};
\end{tikzpicture}
\begin{tikzpicture}[scale=0.7] 
\draw[thick,->] (0,0) -- (6,0) node[anchor=north west] {$y_1$}; 
\draw[thick,->] (0,0) -- (0,6) node[anchor=south east] {$y_2$}; 
\draw[very thick] (0,5) -- (3,0); 
\draw[very thick] (5,0) -- (0,3); 
\draw [fill=black] (1.875,1.875) circle[radius=.1]; 
\draw (0.7,2) node {$x^2_{1|2}$}; \draw (2,0.5) node {$x^2_{2|1}$}; \draw (1.4,4) node [color=blue] {$x^2_{1|1}$}; \draw (4,1.2) node {$x^2_{2|2}$}; \draw (3,-0.5) node {$B^2_{1}$}; \draw (-0.5,3) node {$B^2_{2}$};
\end{tikzpicture}
\par\end{centering}
\caption{Simple-setup for $K=2$ goods and no intersection patches. \label{fig:simple-setup}}
\end{figure}

In this section, we illustrate our setup and Theorem~\ref{thm:main} in the environment with $\mathcal{T}=\{1,2\}$. Consider the setting with two budgets in each time period $B^1_{1},B^1_{2}$ and $B^2_{1},B^2_{2}$ such that $B^t_{1}\cap B^t_{2}\neq \emptyset$ and $w_{1,t}/p_{1,t,K}>w_{2,t}/p_{2,t,K}$ for all $t\in \mathcal{T}$. To simplify the exposition, we assume that demand is continuous, so the intersection patches are picked with probability zero. Thus, in each time period there are four patches $x^t_{1|1},x^t_{2|1},x^t_{1|2}$, and $x^t_{2|2}$ (see Figure~\ref{fig:simple-setup} for a graphical representation of the case with $K=2$ goods).\footnote{Formally, $B^t_1$ is the budget $x^t_{1|1}=\{y\in B^t_1\::\:p\tr_{2,t}y>w_{2,t}\}$, $x^t_{2|1}=\{y\in B^t_1\::\:p\tr_{2,t}y<w_{2,t}\}$, $x^t_{1|2}=\{y\in B^t_2\::\:p\tr_{1,t}y<w_{1,t}\}$, and $x^t_{2|2}=\{y\in B^t_2\::\:p\tr_{1,t}y>w_{1,t}\}$.} We call choice path configurations implied by these 4 patches the \emph{simple-setup} choice paths. An example of a budget path is $\{2,1\}$ (i.e. $B^1_2$ and $B^2_1$), an example of a choice path in this budget path is $\{x_{1|2}^{1},x_{1|1}^{2}\}$. Conditional on a budget path, the total probability of all possible choice paths is equal to $1$ (i.e., $\rho(\{x^1_{1|2},x^2_{1|1}\})+\rho(\{x^1_{2|2},x^2_{1|1}\})+\rho(\{x^1_{1|2},x^2_{2|1}\})+\rho(\{x^1_{2|2},x^2_{2|1}\})=1$). 
\par 
In this setup, there are $3$ rational demand types per time period that are described in Table~\ref{tab:demandtypes1}.\footnote{The idea of writing demand types on patches was developed in \citet{kitamura2018nonparametric} and we use the convenient notation developed in \citet{im2021non}.} Each demand type $\theta^t_{i,j}$ picks $i$-th patch in budget $B_1^t$ and $j$-th patch in budget $B^t_2$ at time $t$.
\begin{table}[h]
\begin{centering}
\begin{tabular}{|c|c|c|}
\hline 
 Type/Budget & $B^t_{1}$ & $B^t_{2}$\\
\hline 
$\theta^t_{1,1}$ & $x^t_{1|1}$ & $x^t_{1|2}$\\
\hline 
$\theta^t_{1,2}$ & $x^t_{1|1}$ & $x^t_{2|2}$\\
\hline 
$\theta^t_{2,2}$ & $x^t_{2|1}$ & $x^t_{2|2}$\\
\hline 
\end{tabular}
\par\end{centering}
\caption{Choices of 3 rational types in budgets $B^t_1$ and $B^t_2$ at time $t$.}\label{tab:demandtypes1}
\end{table}
\par 
Now we can write down the associated $A$ matrix. In this case, since the demand types correspond to a preference type, a demand profile $(\theta^1_{i,j},\theta^2_{k,l})$ (i.e., $\theta^1_{i,j}$ in first time period and $\theta^2_{k,l}$ in the second one) corresponds to a preference profile over the choice path ($9$ preference profiles). The rows of this matrix correspond to the choice paths ($16$ possibles paths). For readability, we replace $0$ by the symbol $``-''$.      
\begin{table}
\begin{centering}
\scalebox{0.6}{
\begin{tabular}{c!{\vrule width 2pt}c|c|c|c|c|c|c|c|c}
 & $(\theta^1_{1,1},\theta^2_{1,1})$ & $(\theta^1_{1,1},\theta^2_{1,2})$ & $(\theta^1_{1,1},\theta^2_{2,2})$ & $(\theta^1_{1,2},\theta^2_{1,1})$ & $(\theta^1_{1,2},\theta^2_{1,2})$ & $(\theta^1_{1,2},\theta^2_{2,2})$ & $(\theta^1_{2,2},\theta^2_{1,1})$ & $(\theta^1_{2,2},\theta^2_{1,2})$ & $(\theta^1_{2,2},\theta^2_{2,2})$\\
\noalign{\hrule height 2pt}
$\{x^1_{1|1},x^2_{1|1}\}$ & $1$ & $1$ & - & 1 & 1 & - & - & - & -\\
\hline 
$\{x^1_{1|1},x^2_{2|1}\}$ & - & - & 1 & - & - & 1 & - & - & -\\
\hline 
$\{x^1_{1|1},x^2_{1|2}\}$ & $1$ & - & - & 1 & - & - & - & - & -\\
\hline 
$\{x^1_{1|1},x^2_{2|2}\}$ & - & 1 & 1 & - & 1 & 1 & - & - & -\\
\hline 
$\{x^1_{2|1},x^2_{1|1}\}$ & - & - & - & - & - & - & 1 & 1 & -\\
\hline 
$\{x^1_{2|1},x^2_{2|1}\}$ & - & - & - & - & - & - & - & - & 1\\
\hline 
$\{x^1_{2|1},x^2_{1|2}\}$ & - & - & - & - & - & - & 1 & - & -\\
\hline 
$\{x^1_{2|1},x^2_{2|2}\}$ & - & - & - & - & - & - & - & 1 & 1\\
\hline 
$\{x^1_{1|2},x^2_{1|1}\}$ & $1$ & 1 & - & - & - & - & - & - & -\\
\hline 
$\{x^1_{1|2},x^2_{2|1}\}$ & - & - & 1 & - & - & - & - & - & -\\
\hline 
$\{x^1_{1|2},x^2_{1|2}\}$ & 1 & - & - & - & - & - & - & - & -\\
\hline 
$\{x^1_{1|2},x^2_{2|2}\}$ & - & 1 & 1 & - & - & - & - & - & -\\
\hline 
$\{x^1_{2|2},x^2_{1|1}\}$ & - & - & - & 1 & 1 & - & 1 & 1 & -\\
\hline 
$\{x^1_{2|2},x^2_{2|1}\}$ & - & - & - & - & - & 1 & - & - & 1\\
\hline 
$\{x^1_{2|2},x^2_{1|2}\}$ & - & - & - & 1 & - & - & 1 & - & -\\
\hline 
$\{x^1_{2|2},x^2_{2|2}\}$ & - & - & - & - & 1 & 1 & - & 1 & 1
\end{tabular}
}
\par\end{centering}
\caption{The matrix $A$ for 2 time periods with 2 budgets per period. $``-''$ corresponds to $0$.}
\end{table}
\par
Matrix $A$ leads to a system of equations $A\nu=\rho$. To simplify the exposition, we represent this system in Table~\ref{table:conerestrictionsex1}. The entries of the table correspond to $\rho(x_{\rand{i}|\rand{j}})$. For example, the element of the matrix in row $3$ and column $1$ corresponds to the equation $\nu_1+\nu_2=\rho(\{x^1_{1|2},x^2_{1|1}\})$. Note that the left upper block of size 2 by 2 of Table~\ref{table:conerestrictionsex1} corresponds to the budget path $\{B^1_1,B^2_1\}$. Hence, the sum of its elements should be 1. The same applies to the right upper, the left bottom, and the right bottom 2 by 2 blocks. 
\par
\begin{table}
\begin{centering}
\begin{tabular}{c!{\vrule width 2pt}c|c|c|c}
& $x^2_{1|1}$ & $x^2_{2|1}$ & $x^2_{1|2}$ & $x^2_{2|2}$\\
\noalign{\hrule height 2pt}
$x^1_{1|1}$ & $\nu_{1}+\nu_{2}+\nu_{4}+\nu_{5}$ & $\nu_{3}+\nu_{6}$ & $\nu_{1}+\nu_{4}$ & $\nu_{2}+\nu_{3}+\nu_{5}+\nu_{6}$\\
\hline 
$x^1_{2|1}$ & $\nu_{7}+\nu_{8}$ & $v_{9}$ & $\nu_{7}$ & $\nu_{8}+\nu_{9}$\\
\hline 
$x^1_{1|2}$ & $\nu_{1}+\nu_{2}$ & $\nu_{3}$ & $\nu_{1}$ & $\nu_{2}+\nu_{3}$\\
\hline 
$x^1_{2|2}$ & $\nu_{4}+\nu_{5}+\nu_{7}+\nu_{8}$ & $\nu_{6}+\nu_{9}$ & $\nu_{4}+\nu_{7}$ & $\nu_{5}+\nu_{6}+\nu_{8}+\nu_{9}$
\end{tabular}
\par\end{centering}
\caption{Matrix representation of $\rho$ under DRUM\label{table:conerestrictionsex1}}
\end{table}

Note that by Theorem~\ref{thm:main}, the existence of $\nu\in\Real^9_{+}$ that solves the system of equations encoded in Table~\ref{table:conerestrictionsex1} is necessary and sufficient for $\rho$ to be consistent with DRUM. Next, we provide three conditions that fully characterize our simple-setup. In the simple-setup the necessity of these conditions for a $\rho$ to be rationalized by DRUM can be verified directly from Table~\ref{table:conerestrictionsex1}. The necessity of these conditions in the general case will be studied later. 

\begin{definition}[Simple Stability] For the simple-setup, $\rho$ satisfies stability if: (i) $\rho(x^2_{i|j};B^1_k)=\sum_{l\in\{1,2\}}\rho(x^1_{l|k};x^2_{i|j})$ does not depend on $B^1_k$ for all $k,i,j\in \{1,2\}$, and (ii)  $\rho(x^1_{l|k};B^2_j)=\sum_{i\in\{1,2\}}\rho(x^1_{l|k};x^2_{i|j})$ does not depend on $B^2_j$ for all $l,k,j\in \{1,2\}$.
\end{definition}
Stability means that the marginal distribution of choices in $t=2$ does not depend on the budget set in $t=1$, and also that the marginal distribution of choices in $t=1$ does not depend on the budget set in $t=2$.  Under stability, the marginal distribution of choices of consumers will not change due to the budget the consumers faced in the past or the budget the consumers will face in the future.  Recall, we have assumed that the stochastic utility process does not depend on the budgets. This condition is an implication of that assumption. 
\par 
The next condition is an analogous condition to the Weak Axiom of Stochastic Revealed Preference (WASRP) for the static case of RUM. To establish this condition, we need a notion of revealed preference on patches. 
\begin{definition}[Patch Revealed Dominance] We say that patch $x^t_{i|j}$ is revealed dominant to $x^t_{i'|j'}$ or $x^t_{i_t|j_t}\succ^D x^t_{i'_t|j'_t}$  if for some $y\in x^t_{i|j}$ and $z\in x^t_{i'|j'}$ (i) $p_{j,t}y> p_{j,t}z$  and (ii) $p_{j',t}y> p_{j',t}z$.
\end{definition}
Patch revealed dominance is a static notion. We can visualize this ordering in Figure~\ref{fig:simple-setup}, where $x^1_{1|1}\succ^D x^1_{1|2}$, and $x^2_{2|2}\succ^D x^2_{2|1}$. The ordering $\succ^D$ combines the requirement that each element on the dominant patch is chosen when each element on the dominated patch is affordable (i.e., strict revealed preference), and the requirement that each element in the dominated patch is not strictly revealed preferred to the elements in the dominant patch. 
Our next condition requires that $\rho$ is monotone on $\succ^D$ in each time period.

\begin{definition}[Simple Monotonicity] For the simple-setup, we say $\rho$ is (simply) monotone if (i)  $x^1_{l|k}\succ^D x^1_{l'|k'}$ implies $\rho(x^1_{l'|k'},x^2_{i|j})\leq \rho(x^1_{l|k},x^2_{i|j})$; and (ii) $x^2_{l|k}\succ^D x^2_{l'|k'}$ implies $\rho(x^1_{i|j},x^2_{l'|k'})\leq \rho(x^1_{i|j},x^2_{l|k})$, for $i,j,l,k\in\{1,2\}$.
\end{definition}
Simple monotonicity is the generalization of the WASRP for our setup. If we had one time period only, a $\rho$ that is consistent with RUM will satisfy WASRP. In fact, for the case of two budgets WASRP also a sufficient condition \citep{hoderlein2014revealed}.  For $2$ time periods, simple monotonicity is not sufficient for the DRUM rationalizability (see Example~\ref{tab:marginalRumNotDrum}). 
\par 
To define the last behavioral implication of DRUM in the simple-setup, we use $\succ^D$ to define a notion of dominance over choice paths.  
\begin{definition}[Choice Path Revealed Dominance] We say that $x_{\rand{i},\rand{j}}$ is revealed dominant to $x_{\rand{i}',\rand{j}'}$ or $x_{\rand{i},\rand{j}}\succ^{D*} x_{\rand{i}',\rand{j}'}$ if for some $\tau\in \mathcal{T}$ $x^{\tau}_{i_{\tau}|j_{\tau}}\succ^D x^{\tau}_{i'_{\tau}|j'_{\tau}}$, and $x^{t}_{i_{t}|j_{t}}\equiv x^{t}_{i_{t}|j_{t}}$, for $t\in\mathcal{T}\setminus{\{\tau\}}$.\footnote{We let $\equiv$ denote set equivalence.}
\end{definition}
In Figure~\ref{fig:simple-setup}, $\{x^1_{1|1},x^2_{1|1}\}$ is revealed dominant to $\{x^1_{1|1},x^2_{1|2}\}$. Also, $\{x^1_{1|2},x^2_{1|1}\}$ is revealed dominant to $\{x^1_{1|2},x^2_{1|2}\}$. 
\begin{definition}[Simple Intensity Monotonicity] For the simple-setup, $\rho$ is (simply) intense monotonic if $\{x^1_{l|k},x^2_{i|j}\}\succ^{D*}\{x^1_{l'|k'};x^2_{i|j}\}$, and $x^2_{i'|j'}\succ^{D}x^2_{i|j}$ then 
\[
\rho(x^1_{l|k},x^2_{i'|j'})-\rho(x^1_{l|k},x^2_{i|j})\geq\rho(x^1_{l'|k'},x^2_{i'|j'})-\rho(x^1_{l'|k'},x^2_{i|j}),
\]
for $i,j,i',j',l,k,l',k'\in\{1,2\}$.
\end{definition}
Simple intensity monotonicity implies that improving a dominant choice path has higher impact on the probability of choice, than improving in the same way a dominated choice path. This property implies that $\rho$ captures intensity of preferences. More formally, we can define an intensity relation.
\begin{definition}[Choice Path Revealed Intensity] We say that 
\[
(\{x^1_{l|k},x^2_{i'|j'}\},\{x^1_{l|k},x^2_{i|j}\})\succ^{I*}(\{x^1_{l'|k'},x^2_{i'|j'}\},\{x^1_{l|k},x^2_{i|j}\}),
\]
if $\{x^1_{l|k},x^2_{i|j}\}\succ^{D*}\{x^1_{l'|k'};x^2_{i|j}\}$, and $x^2_{i'|j'}\succ^{D}x^2_{i|j}$. 
\end{definition}
The ordering $\succ^{I*}$ can be interpreted as $\{x^1_{l|k},x^2_{i'|j'}\}$ dominates $\{x^1_{l|k},x^2_{i|j}\}$ at least as much as $\{x^1_{l'|k'},x^2_{i'|j'}\}$ dominates $\{x^1_{l|k},x^2_{i|j}\}$. Simple intensity monotonicity implies that the difference in $\rho$ between two choice paths, that measures the impact on the probability of choice of switching paths, (i.e., $\rho(x^1_{l|k},x^2_{i'|j'})-\rho(x^1_{l|k},x^2_{i|j})$) is monotone on the intensity relation $\succ^{I*}$. 
We are ready to state our main result in this section.
\begin{theorem}\label{thm:DRUM2x2} 
For the simple-setup, the following are equivalent:
\begin{enumerate}
    \item $\rho$ is rationalized by DRUM. 
    \item $\rho$ satisfies simple stability, simple monotonicity, and simple intensity monotonicity. 
\end{enumerate}
\end{theorem}

Necessity is easy to verify. Sufficiency, is proved constructively. Theorem~\ref{thm:DRUM2x2} is not just a restatement of the Weyl-Minkowski Theorem, as conditions stability and DCLD correspond to the explicit H-representation of the cone restrictions in Table~\ref{table:conerestrictionsex1} (V-representation). The H-representation is obtained by direct computation and can be directly used for testing DRUM in the minimal choice setup. This characterization also provides the reader with a helpful intuition about the empirical content of DRUM.
\par
Next we provide two examples that violate stability, simple monotonicity, or simple intensity monotonicity (i.e., these examples cannot be rationalized by DRUM).  
\begin{example}[Violation of monotonicity and intensity monotonicity] 
Consider the stochastic demand presented in Table~\ref{table:mon+intensity}. It satisfies simple stability. However, it fails to satisfy simple monotonicity and simple intensity monotonicity because $\rho(x^1_{1|2};x^2_{1|2})-\rho(x^1_{1|2};x^2_{1|1})=-\frac{2}{4}$ and $\rho(x^1_{1|1};x^2_{1|2})-\rho(x^1_{1|1};x^2_{1|1})=0$.  
\end{example}

\begin{table}
\begin{centering}
\begin{tabular}{c!{\vrule width 2pt}c|c|c|c}
& $x^2_{1|1}$ & $x^2_{2|1}$ & $x^2_{1|2}$ & $x^2_{2|2}$\\
\noalign{\hrule height 2pt}
$x^1_{1|1}$ & 3/4 & - & 3/4 & -\\
\hline 
$x^1_{2|1}$ & - & 1/4 & 1/4 & -\\
\hline 
$x^1_{1|2}$ & - & 1/4 & 1/4 & -\\
\hline 
$x^1_{2|2}$ & 3/4 & - & 3/4 & -
\end{tabular}
\par\end{centering}
\caption{Matrix representation of $\rho$ that violates simple monotonicity and intensity monotonicity, but satisfies simple stability.}\label{table:mon+intensity}
\end{table}
\par 
Another example of a stochastic demand that fails all 3 conditions of the simple setup is discussed in Section~\ref{sec: unification}.
\section{General Case: Computationally Simple Testable Implications of DRUM \label{sec:generalcasenecessity}}
In this section, we go back to the general primitive and study the necessity of suitable generalizations of the restrictions on behavior introduced in the simple-setup for consistency with DRUM.   First, we define some preliminaries. Let $x^{-t}_{\rand{i}|\rand{j}}=\{x^\tau_{i_{\tau}|j_{\tau}}\}_{\tau\in\mathcal{T}\setminus{t}}$ be the choice path without $t$-th entry. We will abuse notation at treat $x^{-t}_{\rand{i}|\rand{j}}$ as an ordered set and $\cup$ as the operation that inserts an element to it at a $t$-th position. For example, $\{x^{t}_{i_t|j_t}\}\cup x^{-t}_{\rand{i}|\rand{j}}$ denotes $x_{\rand{i}|\rand{j}}=\{x^\tau_{i_{\tau}|j_{\tau}}\}_{\tau\in\mathcal{T}}$. We will also extend the definition of $\rho$ to unions of patches at a given period. In particular, for any $\bigcup_{i\in \mathcal{I}'} x^t_{i|j}$, we define the probability of observing something picked from a choice path $x_{\rand{i}|\rand{j}}$ with $t$-th component replaced by  $\bigcup_{i\in \mathcal{I}'} x^t_{i|j}$ as 
\[
\rho\left(\left\{\bigcup_{i\in \mathcal{I}'} x^t_{i|j}\right\}\cup x^{-t}_{\rand{i}|\rand{j}}\right)=\sum_{i\in\mathcal{I}'}\rho\left(\left\{x^t_{i|j}\right\}\cup x^{-t}_{\rand{i}|\rand{j}}\right).
\]
We present a generalization of the simple stability condition from Section~\ref{sec: 2x2 case}.
\begin{definition}[Stability]
We say that $\rho$ is stable when for any $t\in\mathcal{T}$ and $x^{-t}_{\rand{i}|\rand{j}}$ 
\[
\rho\left(\left\{\bigcup_{i\in \mathcal{I}_{j}^{t}}x^t_{i|j}\right\}\cup x^{-t}_{\rand{i}|\rand{j}}\right)
\]
is the same for all $j\in\mathcal{J}^t$.
\end{definition}
Next, for a given $x^t_{i|j}$ at time $t$ in budget $j$ and any other budget $j'$ define two sets of patches
\begin{align*}
U_{j'}\left(x^t_{i|j}\right)&=\left\{i'\in\mathcal{I}^t_{j'} \::\: x^t_{i'|j'}\succ_D x^t_{i|j}\right\}; &
L_{j'}\left(x^t_{i|j}\right)&=\left\{i'\in\mathcal{I}^t_{j'} \::\: x^t_{i|j}\succ_D x^t_{i'|j'}\right\}.
\end{align*}
and corresponding demands (unions of patches)
\begin{align*}
\rand{x}^t_{1|j,j'}&=\bigcup_{i'\in U_{j'}\left(x^t_{i|j}\right)}x^t_{i'|j'}; &
\rand{x}^t_{2|j,j'}&=\bigcup_{i'\in L_{j'}\left(x^t_{i|j}\right)}x^t_{i'|j'}.
\end{align*}
The sets $U_{j'}\left(x^t_{i|j}\right)$ and $L_{j'}\left(x^t_{i|j}\right)$ capture all the patches in budget $j'$ that dominate and are dominated by $x^t_{i|j}$. The sets $\rand{x}^t_{1|j,j'}$ and $\rand{x}^t_{2|j,j'}$ are just unions of those dominating and dominant patches. 
\par
Now we can state the generalizations of the simple monotonicity and simple intensity monotonicity.
\begin{definition}[Monotonicity]
We say that $\rho$ is monotone when for any $t\in\mathcal{T}$ and $x^{-t}_{\rand{i}|\rand{j}}$ if $x^t_{l|k}\succ^D x^t_{l'|k'}$, then 
\[
\rho\left(\{\rand{x}^t_{1|k,k'}\}\cup x^{-t}_{\rand{i}|\rand{j}}\right)\geq\rho\left(\{\rand{x}^t_{2|k,k'}\}\cup x^{-t}_{\rand{i}|\rand{j}}\right)
\]
\end{definition}

\begin{definition}[Intensity Monotonicity]
We say that $\rho$ is intense monotone when for any $t,t^*\in\mathcal{T}$ and $x^{-t,-t^{*}}_{\rand{i}|\rand{j}}=\{x^{\tau}_{i_\tau|j_\tau}\}_{\tau\in\mathcal{T}\setminus\{t,t^*\}}$ if $x^t_{l|k}\succ^D x^t_{l'|k'}$ and $x^{t^*}_{i|j}\succ^D x^{t^*}_{i'|j'}$, then 
\begin{align*}
&\rho\left(\{\rand{x}^t_{1|k,k'},\rand{x}^{t^*}_{1|j,j'}\}\cup x^{-t,-t^{*}}_{\rand{i}|\rand{j}}\right)-\rho\left(\{\rand{x}^t_{2|k,k'},\rand{x}^{t^*}_{1|j,j'}\}\cup x^{-t,-t^{*}}_{\rand{i}|\rand{j}}\right)\geq \\
&\rho\left(\{\rand{x}^t_{1|k,k'},\rand{x}^{t^*}_{2|j,j'}\}\cup x^{-t,-t^{*}}_{\rand{i}|\rand{j}}\right)-\rho\left(\{\rand{x}^t_{2|k,k'},\rand{x}^{t^*}_{2|j,j'}\}\cup x^{-t,-t^{*}}_{\rand{i}|\rand{j}}\right).
\end{align*}
\end{definition}

\begin{theorem}\label{thm:generalnecessity}
If $\rho$ is rationalized by DRUM, then stability, monotonicity, and intensity monotonicity are satisfied.
\end{theorem}
Stability, monotonicity and intensity monotonicity are not longer sufficient in the general case. An counterexample can be found in Example~$3.2$ in KS, where for the simple case of one time period where RUM is equivalent to DRUM, monotonicity does not imply all conditions that are implied by consistency with RUM. We must highlight that a general characterization of RUM that is analogous to Theorem~\ref{thm:DRUM2x2} (i.e., H-representation of a $\rho$ consistent with DRUM) for the case of more than $3$ goods and more than $3$ budgets is not known. \citet{stoye2019revealed} explains in more detail the computational difficulties of obtaining the H-representation in general cases. The general result in Theorem~\ref{thm:main} has necessary and sufficient conditions for DRUM consistency (i.e., V-representation of a $\rho$ consistent with DRUM)  but in some cases it will be computationally more convenient to check the conditions in Theorem~\ref{thm:generalnecessity} which will provide a conservative test of DRUM. 

\section{Unification of Afriat's theorem and McFadden-Richter's Theorem}\label{sec: unification}
Using Theorem~\ref{thm:main}, we proceed to study the implications of DRUM for simpler domains than $\rho$. In particular, we study the possibility of slicing the panel of choices from budget paths, for a given time period, to obtain a dataset that is a cross-section of choices such as the one described in \citet{mcfadden1990stochastic,mcfadden2005revealed}. We show that any $\rho$ that is consistent with DRUM can be \emph{sliced} to produce a cross-section that is rationalizable by RUM. Then we show that adding the restriction of constant utilities in time as in the Afriat's framework, DRUM implies that the (deterministic) Strong Axiom of Revealed Preference (SARP) has to hold in time-series. This means that DRUM effectively unifies both the \citet{afriat1967construction} and \citet{mcfadden2005revealed} setup into one. 
\par 
We need some preliminaries to formalize our results.  Define, for a given $\tau\in\mathcal{T}$ and a patch $x^{\tau}_{i_{\tau},j_{\tau}}$, the marginal probability is
\[
\rho(x^{\tau}_{i_{\tau}|j_{\tau}},\{B_{j_{t}}^t\}_{t\in\mathcal{T}\setminus{\{\tau\}}})=\sum_{t\in\mathcal{T}\setminus{\{\tau\}}}\sum_{i_t\in \rand{I}^t_{\rand{j}_t}}\rho(\{x^t_{i_t|j_t}\}_{t\in\mathcal{T}}),
\]
where we assume that $x^{\tau}_{i_{\tau},j_{\tau}}$ belongs to the choice path $\rand{j}$. Note that, trivially, we can consider averaging marginal probabilities over repeated budgets (and patches) across time or across choice paths in case the patch $x^{\tau}_{i_{\tau},j_{\tau}}$ appears in more than one choice path.  It will follow that any result for this simpler slice of the panel will hold for these compounded cases. We also need, a preliminary lemma that echoes simple stability. 
\begin{lemma}\label{lemma:pre-stability}
If $\rho$ is rationalized by DRUM, then for any $\tau\in\mathcal{T}$ and any patch $x^{\tau}_{i_{\tau},j_{\tau}}$, the marginal probability $\rho(x^{\tau}_{i_{\tau},j_{\tau}},\{B_j^t\}_{t\in\mathcal{T}\setminus{\{\tau\}}})$ does not depend on $\{B_j^t\}_{t\in\mathcal{T}\setminus{\{\tau\}}}$.
\end{lemma}
The proof of Lemma~\ref{lemma:pre-stability} is omitted because it is trivial. It will also be established as a byproduct of the next result. Given the lemma, for $\rho$ that is rationalized by DRUM, we will omit the dependence on the budgets and write the marginal $\rho$ as $\rho(x^{\tau}_{i_{\tau}|j_{\tau}})$.
\begin{proposition}\label{prop:DRUMimpliesmarginalRUM}
If $\rho$ is rationalized by DRUM, then for any $t\in\mathcal{T}$ there exists a probability measure over $U$, $\mu^t$, such that  
\[
\rho(x^{t}_{i_{t}|j_{t}})= \int \Char{\argmax_{y\in B^t_{j}}w(y)\in x^t_{i_t|j_t}}d\mu^t(w)
\]
for all $j_t\in\mathcal{J}^t$ and $i_t\in\rand{I}^t_{j_t}$.
\end{proposition}
The result in Proposition~\ref{prop:DRUMimpliesmarginalRUM} means that if $\rho$ is consistent with DRUM then slicing the panel of choices will result in a dataset that is consistent with RUM. In this sense, the empirical implications of DRUM when an analyst has access only to a slice of choices is the same as the empirical implications of RUM. However, consistency of the marginal probabilities does not exhaust the empirical content of DRUM. This is illustrated in Example~\ref{ex:RUMmarginalisnotDRUM}.  

\begin{example}\label{ex:RUMmarginalisnotDRUM}[Marginals are consistent with WASRP but not rationalized by DRUM]
Consider the stochastic demand $\rho$ presented in Table~\ref{tab:marginalRumNotDrum}. This $\rho$ violates simple stability, simple monotonicity and simple intensity monotonicity, so DRUM cannot possibly explain it. At the same time its marginal probabilities are consistent with the WASRP: $\rho(x^1_{2|1},B^2_1)=\frac{1}{2}$, $\rho(x^1_{1|2},B^2_1)=\frac{1}{2}$; and $\rho(x^1_{2|1},B^2_2)=\frac{1}{3}$ and $\rho(x^1_{1|2},B^2_2)=\frac{2}{3}$. This means that each of these marginal probabilities is consistent with RUM.\footnote{Recall that WASRP is the necessary and sufficient condition for marginal probabilities to be rationalized by RUM in the sense of Proposition~\ref{prop:DRUMimpliesmarginalRUM}.}
\end{example}

\begin{table}
\begin{centering}
\begin{tabular}{c!{\vrule width 2pt}c|c|c|c}
& $x^2_{1|1}$ & $x^2_{2|1}$ & $x^2_{1|2}$ & $x^2_{2|2}$\\
\noalign{\hrule height 2pt}
$x^1_{1|1}$ & 1/6 & 1/3 & 2/3 & -\\
\hline 
$x^1_{2|1}$ & 1/3 & 1/6 & 1/6 & 1/6\\
\hline 
$x^1_{1|2}$ & 1/6 & 1/3 & 2/3 & -\\
\hline 
$x^1_{2|2}$ & 1/3 & 1/6 & 1/6 & 1/6
\end{tabular}
\par\end{centering}
\caption{Matrix representation of $\rho$ that is consistent with RUM after slicing, but is not consistent with DRUM}\label{tab:marginalRumNotDrum}
\end{table}
\par 
While \citet{mcfadden1990stochastic} study a cross-section of choices from budgets. In the Afriat's framework, only time series of choices from budgets can be used to test utility maximization. However, it is trivial to observe that DRUM has no testable implications for a time-series. This means that if observe $\rho$ on a single budget path (i.e., a time-series), then there are no testable restrictions of DRUM. (We need at least $2$ observed budget paths to test DRUM.) The reason for this is that in the Afriat's framework there is an additional assumption on the stochastic process, namely that $\mu$ assigns positive probability mass only on utility functions such that $u^t=u^s$ all $t,s\in\mathcal{T}$. We call this restriction \emph{constancy} of the stochastic utility process. Under this restriction, the testable implications of DRUM in a time-series are reestablished. We need some preliminaries to formalize this intuition.
\begin{definition}[Strong Axiom of Revealed Path Dominance, SARPD] For a given $\rho$ and a given $\rand{j}\in\rand{J}$, $\rho_{\rand{j}}=(\rho(x_{\rand{i}|\rand{j}}))_{\rand{i}\in\rand{I}_{\rand{j}}}$ satisfies SARPD if 
\[
\rho(\{x^t_{i_t|j_t}\}_{t\in\mathcal{T}})=0,
\]
whenever there is a sequence of patches with elements in $\{x^t_{i_t|j_t}\}_{t\in\mathcal{T}}$ such that $x^t_{i_t|j_t}\succ^D x^s_{i_s|j_s}\succ^D\cdots\succ^D x^k_{i_k|j_k}$ and $x^k_{i_k|j_k}\succ^Dx^t_{i_t|j_t}$.
\end{definition}
Recall that we defined $\succ^D$ to be patch revealed dominance ordering.

\begin{proposition}\label{prop:constantDRUMimpliesSARP} 
If $\rho$ is rationalized by a constant DRUM (i.e., $\mu$ satisfies constancy), then for any given $\rand{j}\in\rand{J}$, $\rho_{\rand{j}}$ satisfies SARPD.
\end{proposition}
We prove here  Proposition~\ref{prop:constantDRUMimpliesSARP} because of its simplicity and interest. Assume towards contradiction that $\rho$ is rationalized by a constant DRUM and SARPD is violated. Then there is a representative element in each patch $x^{t*}_{i_t|j_t},x^{s*}_{i_s|j_s}$ and some utility type  in $\mathcal{U}$, that is the same for all $t\in\mathcal{T}$ (i.e., $u^s=u^t=u$ for all $t,s\in\mathcal{T}$) with positive measure such that $u(x^{t*}_{i_t|j_t})>u(x^{s*}_{i_s|j_s})$. However, the violations of SARPD implies that  $u(x^{t*}_{i_t|j_t})>u(x^{t*}_{i_t|j_t})$ which is impossible. In simple words, SARPD rules out the possibility that there are some individuals in the population that violate the Strong Axiom of Revealed Preferences (SARP). Yet again, constancy of DRUM is what drives testability in a single budget path or time-series. When constancy is relaxed, we need to obtain cross-sectional variation or more than one budget path to reestablish testability of DRUM. 

\section{Pooling \label{sec:pooling}}
In practice, and in the absence of panel variation,  several years or time periods of choices from budgets are pooled before testing for consistency with RUM \citep{kitamura2018nonparametric,deb2017revealed}. Here we explore a potential pitfall of this practice.  We show that a panel dataset that is consistent with DRUM when pooled may not be consistent with RUM. The spurious rejection of rationality may be driven by the fact that pooling requires us to ignore time labels and imposes the restriction that the distribution of preferences is independent across time. 
\par
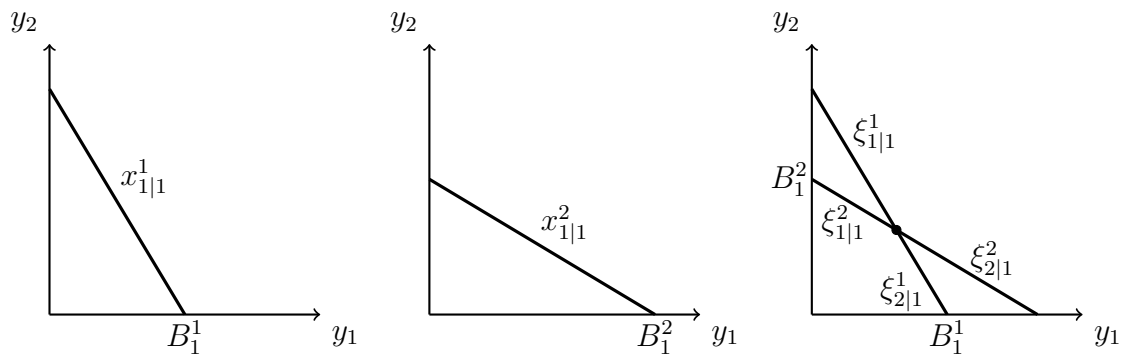
\begin{figure}[h]
\begin{centering}
\begin{tikzpicture}[scale=0.6] 
\draw[thick,->] (0,0) -- (6,0) node[anchor=north west] {$y_1$}; 
\draw[thick,->] (0,0) -- (0,6) node[anchor=south east] {$y_2$}; 
\draw[very thick] (0,5) -- (3,0); 
\draw (2.1,3) node {$x^1_{1|1}$}; 
\draw (3,-0.5) node {$B^1_{1}$}; 
\end{tikzpicture}
\begin{tikzpicture}[scale=0.6] 
\draw[thick,->] (0,0) -- (6,0) node[anchor=north west] {$y_1$}; 
\draw[thick,->] (0,0) -- (0,6) node[anchor=south east] {$y_2$}; 
\draw[very thick] (5,0) -- (0,3); 
\draw (3,2) node {$x^2_{1|1}$}; 
\draw (5,-0.5) node {$B^2_{1}$}; 
\end{tikzpicture}
\begin{tikzpicture}[scale=0.6] 
\draw[thick,->] (0,0) -- (6,0) node[anchor=north west] {$y_1$}; 
\draw[thick,->] (0,0) -- (0,6) node[anchor=south east] {$y_2$}; 
\draw[very thick] (0,5) -- (3,0); 
\draw[very thick] (5,0) -- (0,3); 
\draw [fill=black] (1.875,1.875) circle[radius=.1]; 
\draw (0.7,2) node {$\xi^2_{1|1}$}; 
\draw (2,0.5) node {$\xi^1_{2|1}$}; 
\draw (1.4,4) node {$\xi^1_{1|1}$}; 
\draw (4,1.2) node {$\xi^2_{2|1}$}; 
\draw (3,-0.5) node {$B^1_{1}$}; 
\draw (-0.5,3) node {$B^2_{1}$};
\end{tikzpicture}
\par\end{centering}
\caption{$K=2$ goods, $T=2$ time periods, one budget per time period. The first and the second picture depict patches in 2 time periods. The third picture depicts new patches that arise after pooling the data. \label{fig:pooledpatches}}
\end{figure}

First, we formally define \emph{pooling}. To simplify the exposition, assume that $B_j^t\neq B_{j'}^{t'}$ for all $t,t'\in\mathcal{T}$, $j\in\mathcal{J}^t$, and $j'\in\mathcal{J}^{t'}$. That is, there are no repeated budgets across time and agents. Let $\mathcal{J}=\{1,2,\dots, J\}$, where $J=\sum_{t\in\mathcal{T}}J^t$ is the total number of budgets.
\begin{definition}[Pooled Patches] Let 
\[
\mathcal{X}=\bigcup_{t\in\mathcal{T}}\bigcup_{j\in \mathcal{J}^t} \{\xi^t_{k|j}\}
\]
be the coarsest partition of $\bigcup_{t\in\mathcal{T}}\bigcup_{j\in\mathcal{J}^t}B^t_{j}$ such that
\[
\xi^t_{k|j}\bigcap B^{t}_{j'}\in\{\xi^t_{k|j},\emptyset\}
\]
for any $j,j'$ and $k$.
\end{definition}
The pooled patches $\{\xi^t_{k|j}\}$ partition every $x^t_{i|j}$ since $B^t_{j}$ now may intersect with budgets from different from $t$ time periods (see Figure~\ref{fig:pooledpatches}). 
Given these new patches, we can define the pooled demand $\rho^{\text{pool}}(\xi^t_{k|j})$ as the probability of observing someone picking from patch $\xi^t_{k|j}$. Next we construct a simple example where $\rho$ is rationalizable by DRUM, but the corresponding $\rho^{\text{pool}}$ is not consistent with RUM (in the sense of Proposition~\ref{prop:DRUMimpliesmarginalRUM})
\par
Consider the setting with $K=2$ goods and $T=2$ time periods. In each time period $t$, there is only one budget $B_1^t$. Assume that $B_1^1\neq B_1^2$ and $B_1^1\cup B_1^2\neq\emptyset$ (see Figure~\ref{fig:pooledpatches}). Given that there is on budget variation for any given time period, there is only one choice path $\{x^1_{1|1},x^2_{1|1}\}$. So the trivial $\rho(\{x^1_{1|1},x^2_{1|1}\})=1$ is rationalizable by DRUM. After pooling, since the budgets overlap, there are 4 patches (we assume that the demand is continuous so there is no intersection patches). Since there is only one choice path, DRUM does not impose any restrictions on choice of individuals in these two budgets. As a result, we can take  $\nu^1$ and $\nu^2$ from the DRUM definition such that $\rho^{\text{pool}}(\xi^2_{1|1})+\rho^{\text{pool}}(\xi^1_{2|1})>1$. As a result, this $\rho^{\text{pool}}$ cannot be consistent with RUM.

\section{Conclusion}\label{sec: conclusion}
We have introduced and characterized DRUM, a new model of consumer behavior, when we observe a panel of choices from budget paths. In contrast to the static utility maximization framework, DRUM does not require the assumption that each time period or trial consumers keep their preferences stable over time.
\par 
Our characterization works for any finite collection of choice paths in any finite time window. The characterization can be applied directly in existing panel datasets of consumption using the statistical tools in KS. Our simple-setup characterization showcases that DRUM implies a richer set of behavioral restrictions on the panel of choices than RUM, alleviating some concerns about the empirical bite of the latter in a richer domain.

\bibliographystyle{ecca}
\bibliography{main.bib}

\begin{thebibliography}{32}
\providecommand{\natexlab}[1]{#1}

\bibitem[{Adams \textit{et~al.}(2015)Adams, Blundell, Browning and
  Crawford}]{adams2015prices}
\textsc{Adams, A.}, \textsc{Blundell, R.}, \textsc{Browning, M.} and
  \textsc{Crawford, I.} (2015). \textit{Prices versus preferences: taste change
  and revealed preference}. Tech. rep., IFS Working Papers.

\bibitem[{Afriat(1967)}]{afriat1967construction}
\textsc{Afriat, S.~N.} (1967). The construction of utility functions from
  expenditure data. \textit{International economic review}, \textbf{8}~(1),
  67--77.

\bibitem[{Aguiar and Kashaev(2021)}]{AK2021}
\textsc{Aguiar, V.~H.} and \textsc{Kashaev, N.} (2021). Stochastic revealed
  preferences with measurement error. \textit{The Review of Economic Studies},
  \textbf{88}~(4), 2042--2093.

\bibitem[{Ahn \textit{et~al.}(2014)Ahn, Choi, Gale and
  Kariv}]{ahn2014estimating}
\textsc{Ahn, D.}, \textsc{Choi, S.}, \textsc{Gale, D.} and \textsc{Kariv, S.}
  (2014). Estimating ambiguity aversion in a portfolio choice experiment.
  \textit{Quantitative Economics}, \textbf{5}~(2), 195--223.

\bibitem[{Block and Marschak(1960)}]{block1960random}
\textsc{Block, H.} and \textsc{Marschak, J.} (1960). Random orderings and
  stochastic theories of responses". in i. olkin, s. ghurye, w. hoeffding, w.
  madow, and h. man (eds) contributions to probability and statistics, stanford
  university press.

\bibitem[{Border(2007)}]{border2007introductory}
\textsc{Border, K.} (2007). Introductory notes on stochastic rationality.
  \textit{California Institute of Technology}.

\bibitem[{Browning(1989)}]{browning1989anonparametric}
\textsc{Browning, M.} (1989). A nonparametric test of the life-cycle rational
  expections hypothesis. \textit{International Economic Review}, pp. 979--992.

\bibitem[{Chambers \textit{et~al.}(2021)Chambers, Masatlioglu and
  Turansick}]{chambers2021correlated}
\textsc{Chambers, C.~P.}, \textsc{Masatlioglu, Y.} and \textsc{Turansick, C.}
  (2021). Correlated choice. \textit{arXiv preprint arXiv:2103.05084}.

\bibitem[{Cherchye \textit{et~al.}(2017)Cherchye, Demuynck, De~Rock, Vermeulen
  \textit{et~al.}}]{cherchye2017household}
\textsc{Cherchye, L.}, \textsc{Demuynck, T.}, \textsc{De~Rock, B.},
  \textsc{Vermeulen, F.} \textit{et~al.} (2017). Household consumption when the
  marriage is stable. \textit{American Economic Review}, \textbf{107}~(6),
  1507--1534.

\bibitem[{Choi \textit{et~al.}(2007)Choi, Fisman, Gale and
  Kariv}]{choi2007revealing}
\textsc{Choi, S.}, \textsc{Fisman, R.}, \textsc{Gale, D.~M.} and \textsc{Kariv,
  S.} (2007). Revealing preferences graphically: an old method gets a new tool
  kit. \textit{American Economic Review}, \textbf{97}~(2), 153--158.

\bibitem[{Choi \textit{et~al.}(2014)Choi, Kariv, M{\"u}ller and
  Silverman}]{choi2014more}
\textsc{---}, \textsc{Kariv, S.}, \textsc{M{\"u}ller, W.} and
  \textsc{Silverman, D.} (2014). Who is (more) rational? \textit{The American
  Economic Review}, \textbf{104}~(6), 1518--1550.

\bibitem[{Dean and Martin(2016)}]{dean2016measuring}
\textsc{Dean, M.} and \textsc{Martin, D.} (2016). Measuring rationality with
  the minimum cost of revealed preference violations. \textit{Review of
  Economics and Statistics}, \textbf{98}~(3), 524--534.

\bibitem[{Deb \textit{et~al.}(2017)Deb, Kitamura, Quah and
  Stoye}]{deb2017revealed}
\textsc{Deb, R.}, \textsc{Kitamura, Y.}, \textsc{Quah, J. K.-H.} and
  \textsc{Stoye, J.} (2017). Revealed price preference: Theory and stochastic
  testing. \textit{Working paper}.

\bibitem[{Echenique \textit{et~al.}(2011)Echenique, Lee and
  Shum}]{echenique2011money}
\textsc{Echenique, F.}, \textsc{Lee, S.} and \textsc{Shum, M.} (2011). The
  money pump as a measure of revealed preference violations. \textit{Journal of
  Political Economy}, \textbf{119}~(6), 1201--1223.

\bibitem[{Frick \textit{et~al.}(2019)Frick, Iijima and
  Strzalecki}]{frick2019dynamic}
\textsc{Frick, M.}, \textsc{Iijima, R.} and \textsc{Strzalecki, T.} (2019).
  Dynamic random utility. \textit{Econometrica}, \textbf{87}~(6), 1941--2002.

\bibitem[{Gauthier(2018)}]{gauthier2018}
\textsc{Gauthier, C.} (2018). Nonparametric identification of discount factors
  under partial efficiency. \textit{Working paper}.

\bibitem[{Gauthier(2021)}]{gauthier2021}
\textsc{---} (2021). Price search and consumption inequality: Robust, credible,
  and valid inference. \textit{Working paper}.

\bibitem[{Hoderlein and Stoye(2014)}]{hoderlein2014revealed}
\textsc{Hoderlein, S.} and \textsc{Stoye, J.} (2014). Revealed preferences in a
  heterogeneous population. \textit{Review of Economics and Statistics},
  \textbf{96}~(2), 197--213.

\bibitem[{Im and Rehbeck(2021)}]{im2021non}
\textsc{Im, C.} and \textsc{Rehbeck, J.} (2021). Non-rationalizable
  individuals, stochastic rationalizability, and sampling. \textit{Available at
  SSRN 3767994}.

\bibitem[{Kawaguchi(2017)}]{kawaguchi2017testing}
\textsc{Kawaguchi, K.} (2017). Testing rationality without restricting
  heterogeneity. \textit{Journal of Econometrics}, \textbf{197}~(1), 153--171.

\bibitem[{Kitamura and Stoye(2018)}]{kitamura2018nonparametric}
\textsc{Kitamura, Y.} and \textsc{Stoye, J.} (2018). Nonparametric analysis of
  random utility models. \textit{Econometrica}, \textbf{86}~(6), 1883--1909.

\bibitem[{Kurtz-David \textit{et~al.}(2019)Kurtz-David, Persitz, Webb and
  Levy}]{kurtz2019neural}
\textsc{Kurtz-David, V.}, \textsc{Persitz, D.}, \textsc{Webb, R.} and
  \textsc{Levy, D.~J.} (2019). The neural computation of inconsistent choice
  behavior. \textit{Nature communications}, \textbf{10}~(1), 1583.

\bibitem[{Li(2021)}]{li2021axiomatization}
\textsc{Li, R.} (2021). An axiomatization of stochastic utility. \textit{arXiv
  preprint arXiv:2102.00143}.

\bibitem[{McFadden and Richter(1990)}]{mcfadden1990stochastic}
\textsc{McFadden, D.} and \textsc{Richter, M.~K.} (1990). Stochastic
  rationality and revealed stochastic preference. \textit{Preferences,
  Uncertainty, and Optimality, Essays in Honor of Leo Hurwicz, Westview Press:
  Boulder, CO}, pp. 161--186.

\bibitem[{McFadden(2005)}]{mcfadden2005revealed}
\textsc{McFadden, D.~L.} (2005). Revealed stochastic preference: a synthesis.
  \textit{Economic Theory}, \textbf{26}~(2), 245--264.

\bibitem[{Ok(2011)}]{ok2011real}
\textsc{Ok, E.~A.} (2011). \textit{Real analysis with economic applications}.
  Princeton University Press.

\bibitem[{Polisson \textit{et~al.}(2020)Polisson, Quah and
  Renou}]{polisson2020revealed}
\textsc{Polisson, M.}, \textsc{Quah, J. K.-H.} and \textsc{Renou, L.} (2020).
  Revealed preferences over risk and uncertainty. \textit{American Economic
  Review}, \textbf{110}~(6), 1782--1820.

\bibitem[{Porter and Adams(2016)}]{porter2016love}
\textsc{Porter, M.} and \textsc{Adams, A.} (2016). For love or reward?
  characterising preferences for giving to parents in an experimental setting.
  \textit{The Economic Journal}, \textbf{126}~(598), 2424--2445.

\bibitem[{Ray and Robson(2018)}]{ray2018certified}
\textsc{Ray, D.} and \textsc{Robson, A.} (2018). Certified random: A new order
  for coauthorship. \textit{American Economic Review}, \textbf{108}~(2),
  489--520.

\bibitem[{Smeulders \textit{et~al.}(2021)Smeulders, Cherchye and
  De~Rock}]{smeulders2021nonparametric}
\textsc{Smeulders, B.}, \textsc{Cherchye, L.} and \textsc{De~Rock, B.} (2021).
  Nonparametric analysis of random utility models: computational tools for
  statistical testing. \textit{Econometrica}, \textbf{89}~(1), 437--455.

\bibitem[{Stoye(2019)}]{stoye2019revealed}
\textsc{Stoye, J.} (2019). Revealed stochastic preference: A one-paragraph
  proof and generalization. \textit{Economics Letters}, \textbf{177}, 66--68.

\bibitem[{Strzalecki(2021)}]{straleckinotes}
\textsc{Strzalecki, T.} (2021). \textit{Stochastic Choice}. Mimeo.

\end{thebibliography}

\section{Proofs}\label{app: proofs}
\subsection{Proof of Theorem~\ref{thm:main}}
($(i)\iff(ii)\iff(iii)$)
\par
In this proof, we adapt the proof Theorem~$3.1$ in KS for RUM for the dynamic case.  Our proof uses profiles of nonstochastic demand profiles. 
For each time period $t\in \mathcal{T}$ we define the nonstochastic demand types as in KS: $(\theta^t_{1},\cdots,\theta^t_{J^t})\in B^t_{1}\times\cdots\times B^t_{J^t}$. This system of types is rationalizable if $\theta^t_{j}\in \argmax_{y\in B^t_{j}}u^t(y)$ for $j=1,\cdots,J^t$ for some utility function $u^t$. 
\par
Then we form any given nonstochastic demand profile by stacking up the demand types in a budget path $\rand{j}$ as $\theta_{\rand{j}}=(\theta^t_{j_t})_{j_t\in \rand{j}}$.  
\par 
Fix $\rho$. For fixed $t\in\mathcal{T}$, let the set $\mathcal{Y}^*_t$ collect the geometric center point of each patch. Let $\rho^*$ be the unique dynamic stochastic demand system concentrated on $\mathcal{Y}^*_t$ for all $t\in \mathcal{T}$. KS established that demand systems can be arbitrarily perturbed  within patches in a given time period $t$. such that $\rho$ is rationalizable by DRUM if and only if $\rho^*$ is. It follows that the rationalizability of $\rho$ can be decided by checking whether there exists a mixture of nonstochastic demand profiles supported on $\mathcal{Y}^*_t$ for all $t\in\mathcal{T}$.
\par 
Since we have assumed a finite number of budgets, and time periods, there will be a finite number of budget paths, using our notation we have $\abs{\rand{J}}$ budget paths. Also, because $\mathcal{Y}^*_t$ is finite for all $t\in\mathcal{T}$, there are finitely many nonstochastic demand profiles. Noting that these demand profiles are characterized by binary vector representation corresponding to columns of $A$, the statement of the theorem follows immediately.  
\par
($(i)\iff(iv)$)
\par
$(i)\implies (iv)$ is trivial. 
\par 
The proof $(iv) \implies (i)$ is completely analogous to the proof for the case of RUM in \citet{border2007introductory}. We just need to replace the system of equations in that proof with the one we describe in Theorem~\ref{thm:main}.(ii). The rest of the proof follows from Farkas' lemma.

\subsection{Proof of Theorem~\ref{thm:DRUM2x2}}
\textbf{Necessity.} Suppose there is $\nu\in\Real^{9}_{+}$ that solves the system. That is
\begin{align}
    \rho_1&=\nu_1+\nu_2+\nu_4+\nu_5\\
    \rho_2&=\nu_3+\nu_6\\
    \rho_3&=\nu_1+\nu_4\\
    \rho_4&=\nu_2+\nu_3+\nu_5+\nu_6\\
    \rho_5&=\nu_7+\nu_8\\
    \rho_6&=\nu_9\\
    \rho_7&=\nu_7\\
    \rho_8&=\nu_8+\nu_9\\
    \rho_9&=\nu_1+\nu_2\\
    \rho_{10}&=\nu_3\\
    \rho_{11}&=\nu_1\\
    \rho_{12}&=\nu_2+\nu_3\\
    \rho_{13}&=\nu_4+\nu_5+\nu_7+\nu_8\\
    \rho_{14}&=\nu_6+\nu_9\\
    \rho_{15}&=\nu_4+\nu_7\\
    \rho_{16}&=\nu_5+\nu_6+\nu_8+\nu_9.
\end{align}
Thus, the system can be split in 3 blocks. Block 1 is derived from equations (6), (7), (10), and (11):
\begin{align*}
    \nu_1&=\rho_{11}\\
    \nu_3&=\rho_{10}\\
    \nu_7&=\rho_7\\
    \nu_9&=\rho_6.
\end{align*}
Block 2 is derived using Block 1 and equations (2), (3), (5), (8), (9), (12), (14), and (15):
\begin{align*}    
    \nu_2&=\rho_9-\rho_{11}=\rho_{12}-\rho_{10}\\
    \nu_4&=\rho_{15}-\rho_7=\rho_3-\rho_{11}\\
    \nu_6&=\rho_{14}-\rho_6=\rho_2-\rho_{10}\\
    \nu_8&=\rho_5-\rho_7=\rho_8-\rho_6.
\end{align*}
Block 3 is derived from Block 1 and equations (1), (4), (13), and (16):
\begin{align*}
    \nu_5&=\rho_{16}-\rho_{6}-(\nu_6+\nu_8)=\rho_{13}-\rho_{7}-(\nu_4+\nu_8)=\rho_4-\rho_{10}-(\nu_2+\nu_6)=\rho_1-\rho_{11}-(\nu_2+\nu_4).
\end{align*}
The simplex restriction on $\rho$ (e.g., $\rho_1+\rho_2+\rho_5+\rho_6=1$) and Block 2 imply simple stability. The fact that $v\in\Real^{9}_{+}$ and Block 2 and Block 3 equations imply simple monotonicity and simple intensity monotonicity: simple monotonicity follows from Block 2 equations, simple intensity monotonicity follows from Block 2 and Block 3 equations since
\[
0 \leq \nu_5=\rho_1-\rho_{11}-(\nu_2+\nu_4)=\rho_1-\rho_{11}-\rho_9+\rho_{11}-\rho_3+\rho_{11}=\rho_1-\rho_9-\rho_3+\rho_{11}.
\]
\par
\textbf{Sufficiency.} Suppose $\rho$ satisfies simple stability, simple monotonicity, and simple intensity monotonicity. Consider the following $v\in\Real^9$:
\begin{align*}
    \nu_1&=\rho_{11} & \nu_4&=\rho_3-\rho_{11} & \nu_7&=\rho_7\\
    \nu_2&=\rho_9-\rho_{11} & \nu_5&=\rho_1-\rho_9-\rho_3+\rho_{11} & \nu_8&=\rho_5-\rho_7\\
    \nu_3&=\rho_{10} & \nu_6&=\rho_2-\rho_{10} & \nu_9&=\rho_6.
\end{align*}
Nonnegativity of $\nu$ follows from nonnegativity of $\rho$ and stability combined with simple monotonicity and simple intensity monotonicity (e.g. simple intensity monotonicity implies $\nu_5\geq 0$). 
\par
It is left to show that the proposed $\nu$ solves system (1)-(16). Equations (1)-(3), (5)-(7), (9)-(11) are trivially satisfied. Equation (4) is satisfied since
\begin{align*}
    \nu_2+\nu_3+\nu_5+\nu_6=\rho_1+\rho_2-\rho_3=\rho_3+\rho_4-\rho_3=\rho_4,
\end{align*}
where the second equality follows from stability. Similarly, stability implies equations (8), (12), (13), (14), and (15):
\begin{align*}
    \nu_8+\nu_9=\rho_5+\rho_6-\rho_7=\rho_7+\rho_8-\rho_7=\rho_8;\\
    \nu_2+\nu_3=\rho_9+\rho_{10}-\rho_{11}=\rho_{11}+\rho_{12}-\rho_{11}=\rho_{12};\\
    \nu_4+\nu_5+\nu_7+\nu_8=\rho_1+\rho_5-\rho_9=\rho_9+\rho_{13}-\rho_9=\rho_{13};\\
    \nu_6+\nu_9=\rho_2+\rho_6-\rho_{10}=\rho_{10}+\rho_{14}-\rho_{10}=\rho_{14};\\
    \nu_4+\nu_7=\rho_3+\rho_7-\rho_{11}=\rho_{11}+\rho_{15}-\rho_{11}=\rho_{15}.
\end{align*}
Equation (16) is satisfied since
\begin{align*}
    &\nu_5+\nu_6+\nu_8+\nu_9=\rho_1-\rho_9-\rho_3+\rho_{11}+\rho_2-\rho_{10}+\rho_5-\rho_7+\rho_6=\\
    &1-(\rho_3+\rho_7)-(\rho_9+\rho_{10})+\rho_{11}=1-(\rho_{11}+\rho_{15})-(\rho_{11}+\rho_{12})+\rho_{11}\\
    &1-\rho_{11}-\rho_{12}-\rho_{15}=\rho_{16},
\end{align*}
where the second and the last equalities follow from the simplex restriction on $\rho$ (e.g., $\rho_1+\rho_2+\rho_5+\rho_6=1$); and the third equality follows from stability.

\subsection{Proof of Proposition~\ref{prop:DRUMimpliesmarginalRUM}}
By the definition of rationalizability by DRUM
\[
\rho(\{x^t_{i_t|j_t}\}_{t\in \mathcal{T}})=\int \prod_{t\in \mathcal{T}} \Char{\argmax_{y\in B^t_{j}}u^t(y)\in x^t_{i_t|j_t}}d\mu(u)\quad,\forall \rand{i},\rand{j},
\]
for some measure $\mu$. 
Now we compute
\[
\rho(\{x^t_{i_t|j_t}\}_{t\in\mathcal{T}\setminus{\{\tau\}}},B^{\tau}_{j})=\sum_{i_{\tau}\in \rand{I}^{\tau}_{\rand{j}_{\tau}}}\rho(\{x^t_{i_t|j_t}\}_{t\in\mathcal{T}}).
\]
This means that
\[
\rho(\{x^t_{i_t|j_t}\}_{t\in\mathcal{T}\setminus{\{\tau\}}},B^{\tau}_{j})=\sum_{i_{\tau}\in\rand{I}^{\tau}_{\rand{j}_{\tau}}}(\int \prod_{t\in \mathcal{T}}\Char{\argmax_{y\in B^t_{j}}u^t(y)\in x^t_{i_t|j_t}}d\mu(u)).
\]
In the RHS of the previous equation, we can exchange the external summation operator with the integral and obtain:
\[
\rho(\{x^t_{i_t|j_t}\}_{t\in\mathcal{T}\setminus{\{\tau\}}},B^{\tau}_{j})=(\int \sum_{i_{\tau}\in\rand{I}^{\tau}_{\rand{j}_{\tau}}}\prod_{t\in \mathcal{T}}\Char{\argmax_{y\in B^t_{j}}u^t(y)\in x^t_{i_t|j_t}}d\mu(u)).
\]
By \citet{ok2011real} we know that if we maximize a continuous and monotone utility function subject to a linear budget constraint there is always a maximum on the budget line. This implies that   
\[
\sum_{i_{\tau}\in\rand{I}^{\tau}_{\rand{j}_{\tau}}}\prod_{t\in \mathcal{T}}\Char{\argmax_{y\in B^t_{j_t}}u^t(y)\in x_{i_t|j_t}}=\prod_{t\in \mathcal{T}\setminus{\{\tau\}}}\Char{\argmax_{y\in B^t_{j_t}}u^t(y)\in x_{i_t|j_t}}.
\]
\par 
If we iterate this step, then we conclude that 
\[
\sum_{t\in\mathcal{T}\setminus{\{\tau\}}}\sum_{i_t\in\rand{I}^{\tau}_{\rand{j}_{\tau}}}\rho(\{x^t_{i_t|j_t}\}_{t\in\mathcal{T}})=\int\Char{\argmax_{y\in B^{\tau}_{j}}u^{\tau}(y)\in x^{\tau}_{i_{\tau}|j_{\tau}}}d\mu(u).
\]

\section{Proof of Theorem~\ref{thm:generalnecessity}}

\textbf{Stability.} By the definition of  DRUM, there exists a distribution over $\mathcal{U}$, $\mu$, such that
\[
\rho\left(\{x_{i_t|j_t}\}_{t\in \mathcal{T}}\right)=\int \prod_{t\in \mathcal{T}} \Char{\argmax_{y\in B^t_{j_t}}u^t(y)\in x^t_{i_t|j_t}}d\mu(u)
\]
for all $\rand{i},\rand{j}$. Fix some $t\in\mathcal{T}$ and $x^{-t'}_{\rand{i}|\rand{j}}$ and $j\in\mathcal{J}^{t'}$ and note that 
\begin{align*}
&\rho\left(\left\{\bigcup_{i\in \mathcal{I}_{j}^{t'}}x^{t'}_{i|j}\right\}\cup x^{-t'}_{\rand{i}|\rand{j}}\right)=\\
&\sum_{i\in \mathcal{I}_{j}^{t'}} \int \prod_{t\in \mathcal{T}\setminus\{t'\}} \Char{\argmax_{y\in B^t_{j_t}}u^t(y)\in x^t_{i_t|j_t}}\Char{\argmax_{y\in B^{t'}_{j}}u^{t'}(y)\in x^{t'}_{i|j}}d\mu(u)=\\
&\int \prod_{t\in \mathcal{T}\setminus\{t'\}} \Char{\argmax_{y\in B^t_{j_t}}u^t(y)\in x^t_{i_t|j_t}}\sum_{i\in \mathcal{I}_{j}^{t'}} \Char{\argmax_{y\in B^{t'}_{j}}u^{t'}(y)\in x^{t'}_{i|j}}d\mu(u)=\\
&\int \prod_{t\in \mathcal{T}\setminus\{t'\}} \Char{\argmax_{y\in B^t_{j_t}}u^t(y)\in x^t_{i_t|j_t}}d\mu(u),
\end{align*}
where the last equality follows from $\argmax_{y\in B^{t'}_{j_{t'}}}u^{t'}(y)$ being a singleton ($u^{t'}$ is continuous and strictly monotone) and $\{x^{t'}_{i'|j_{t'}}\}_{i\in\mathcal{I}_{j_{t'}}^{t'}}$ being a partition. The right-hand side of the last expression does not depend on the choice of $j$. Stability follows since the choice of $t'$ and $x^{-t'}_{\rand{i}|\rand{j}}$ was arbitrary.

\textbf{Monotonicity and Intensity Monotonicity.} Fix any $t,t^*\in\mathcal{T}$; any two intersecting budgets in each time period, $k,k'\in\mathcal{J}^t$ and $j,j'\in\mathcal{J}^{t^*}$ period, and $x^{-t,-t^*}_{\rand{i}|\rand{j}}$. Note that conditional on choices in all time periods but $t,t^*$ being $x^{-t,-t^*}_{\rand{i}|\rand{j}}$, the problem is described by the simple-setup choice paths (2 time periods and 2 budgets). Next since $\rho$ is rationalizable by DRUM, then the conditional probability over the simple-setup choice path is also rationalizable by DRUM. Hence, the conditional probabilities should satisfy simple monotonicity and simple intensity monotonicity. Multiplying all inequalities that define simple monotonicity and simple intensity monotonicity by the probability of observing $x^{-t,-t^*}_{\rand{i}|\rand{j}}$ deliver the inequalities implied by monotonicity and intensity monotonicity. The fact that the choice of $t,t',k,k',j,j'$, and $x^{-t,-t^*}_{\rand{i}|\rand{j}}$ was arbitrary completes the proof.

\end{document}